\documentclass[epj]{webofc}
\usepackage[varg]{txfonts}   
\woctitle{MESON2018 - the 15$^\textrm{th}$ International Workshop on Meson Physics}
\begin{document}
\selectlanguage{english}
\title{Importance of mesons in light-by-light scattering \\ in ultraperipheral lead-lead collisions at the LHC}

\author{Mariola K{\l}usek-Gawenda\inst{1}\fnsep\thanks{\email{mariola.klusek@ifj.edu.pl}}
}

\institute{Institute of Nuclear Physics Polish Academy of Sciences, PL-31-342 Krakow, Poland}

\abstract{
  Possibility of PbPb$\to$PbPb$\gamma\gamma$ measurement
  at smaller (< 5 GeV) diphoton invariant mass will be presented.
  Analysis focuses only on ultraperipheral heavy-ion collisions.
  This estimate shows that the $\gamma \gamma \to \gamma \gamma$ collisions 
  can be measured at the LHC by ALICE and LHCb experiments
  for diphoton invariant mass $>$ 2 GeV. 
  Predictions for the $\gamma\gamma \to \eta, \eta' \to \gamma\gamma$
  resonance scattering shows that these resonances 
  can be measured with rather good statistics.
  A~possible observation of peaks 
  related to intermediate $\eta$, $\eta'(958)$, $\eta_c(1S)$, $\eta_c(2S)$, $\chi_{c0}(1P)$ mesons will be presented too.
  Attempts of reduction of background which comes from dipion production 
  will be considered.}
\maketitle
\section{Introduction}
\label{intro}

Light-by-light scattering was realized experimentally only recently
\cite{Aaboud:2017bwk, CMS_LbL} in ultraperipheral ultrarelativistic heavy ion
collisions. 
For ions of charges $Z_1$, $Z_2$, the cross
section is enhanced by $Z_1^2 Z_2^2$ factor compared to
proton-proton collisions, at least at low diphoton
invariant masses equal to diphoton collision energies,
where the initial photons are quasi real with extremely low virtualities.
ATLAS measured a fiducial cross section of $\sigma = 70 \pm 24$ (stat.) $\pm 17$
(syst.)~nb and our theoretical calculations (including experimental acceptance) gave 
$49 \pm 10$~nb \cite{Klusek-Gawenda:2016euz}. 
ATLAS comparison of its experimental results to the predictions
from Ref.\cite{Klusek-Gawenda:2016euz} show a reasonable agreement. 
In comparison,
the CMS Collaboration measured the same process but for smaller $\gamma\gamma$ 
invariant mass ($M_{\gamma\gamma}$ > 5 GeV).
Our theoretical calculations $103 \pm 0.034$~nb are in the good agreement
with the experimental fiducial cross section: $\sigma = 122 \pm 46$ (stat.) $\pm 29$ (syst.)~nb.

This note will answer to a question
whether we can go to lower $\gamma\gamma$ scattering energies at the LHC.
The calculations are directed to ALICE and LHCb experimental limitation.
At~lower energies ($W_{\gamma\gamma} <$ 4 GeV) meson 
resonances may play some role in addition to the Standard Model box 
diagrams \cite{Lebiedowicz:2017cuq}
or double photon fluctuations into light vector mesons \cite{Klusek-Gawenda:2016euz}
or two-gluon exchanges \cite{Klusek-Gawenda:2016nuo}.

\section{Theory}
\label{sec-1}

The phase space integrated cross section for nuclear ultraperipheral collisions (UPCs) 
in impact parameter space ($b$) can be expressed through the five-fold integral

\begin{eqnarray}
\sigma_{PbPb \to Pb Pb X_1 X_2}\left(\sqrt{s_{PbPb}} \right) &=&
\int \sigma_{\gamma \gamma \to X_1 X_2} 
\left(W_{\gamma\gamma} \right)
N\left(\omega_1, {\bf b_1} \right)
N\left(\omega_2, {\bf b_2} \right)  \, S_{abs}^2\left({\bf b}\right) \nonumber  \\ 
& \times &
\mathrm{d}^2 b \, \mathrm{d}\overline{b}_x \, \mathrm{d}\overline{b}_y \, 
\frac{W_{\gamma\gamma}}{2}
\mathrm{d} W_{\gamma\gamma} \, \mathrm{d} Y_{X_1 X_2} \;,
\label{eq:EPA_sigma_final_5int}
\end{eqnarray} 
where here $X_1X_2$ is a pair of photons or neutral pions.
Incoming photons energy
$W_{\gamma\gamma}=\sqrt{4\omega_1\omega_2}$ is expressed through
energy of the single photon $\omega_i$
and $Y_{X_1 X_2}=\left( y_{X_1} + y_{X_2} \right)/2$ 
is rapidity of the outgoing system. 
$\bf b_1$ and $\bf b_2$ are impact parameters 
of the photon-photon collision point with respect to parent
nuclei 1 and 2, respectively, 
and ${\bf b} = {\bf b_1} - {\bf b_2}$ is the standard impact parameter 
for the $A_1 A_2$ collision.
The absorption factor $S_{abs}^2\left({\bf b}\right)$ assures UPC 
which means that the nuclei do not undergo nuclear breakup.
The photon flux ($N\left(\omega_i, {\bf b_i} \right)$) is expressed through a nuclear charge form factor of the nucleus. In our calculations we use a~realistic 
form factor which is a Fourier transform of the charge distribution 
in the nucleus. More details can be found e.g. in \cite{KlusekGawenda:2010kx}.
The elementary cross section $\sigma_{\gamma \gamma \to X_1 X_2}$ in 
Eq.~(\ref{eq:EPA_sigma_final_5int}) for
the $\gamma \gamma \to \gamma \gamma$ scattering is calculated 
within LO QED with fermion loops (see left panel of Fig.~1 in Ref.~\cite{Klusek-Gawenda:2016euz}).
The elementary cross section for $\gamma \gamma \to \pi \pi$ was studied in detail
in Ref.~\cite{Klusek-Gawenda:2013rtu}.
\begin{figure}[h]
	\centering
	(a)\includegraphics[scale=0.22]{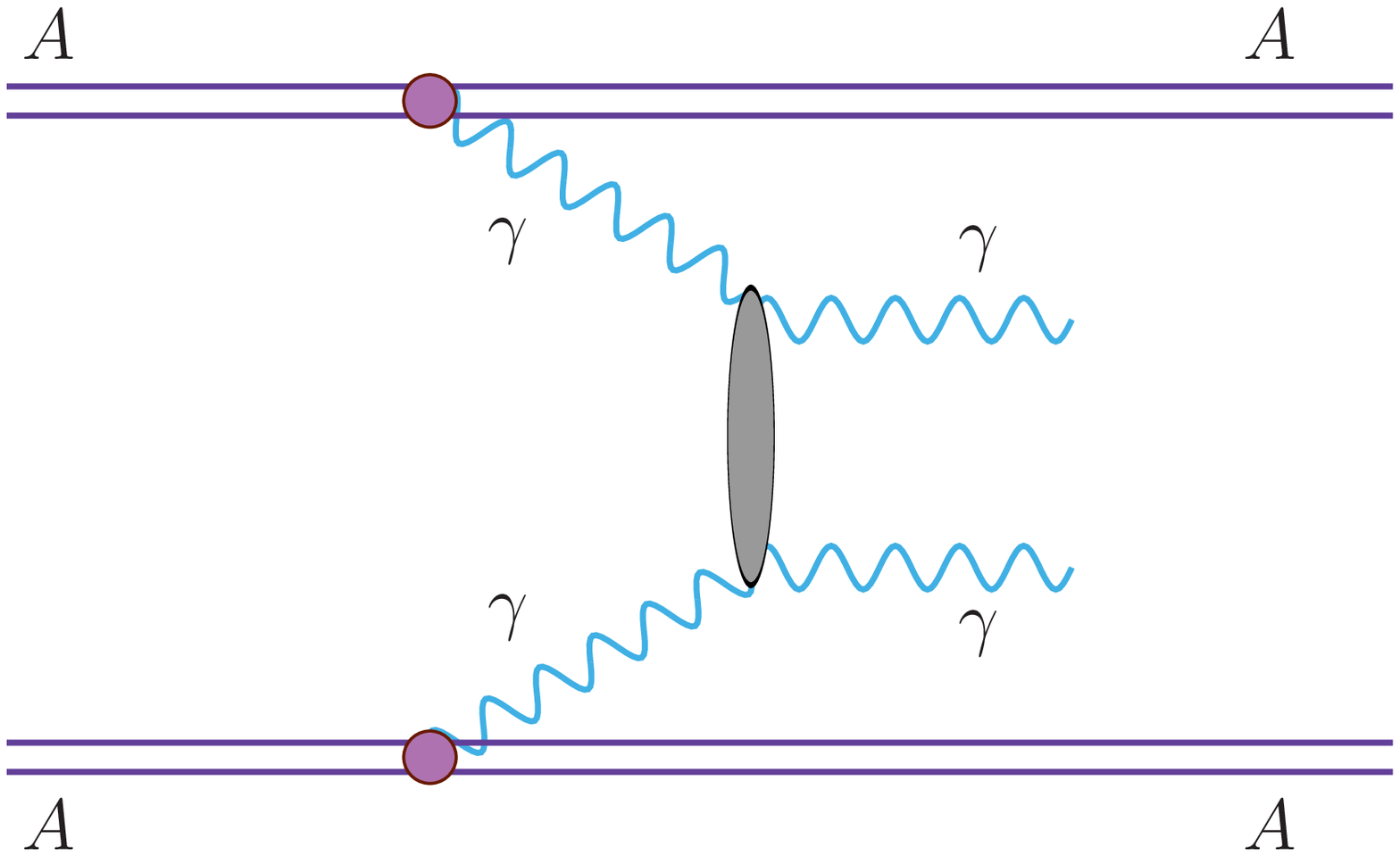} 
	(b)\includegraphics[scale=0.22]{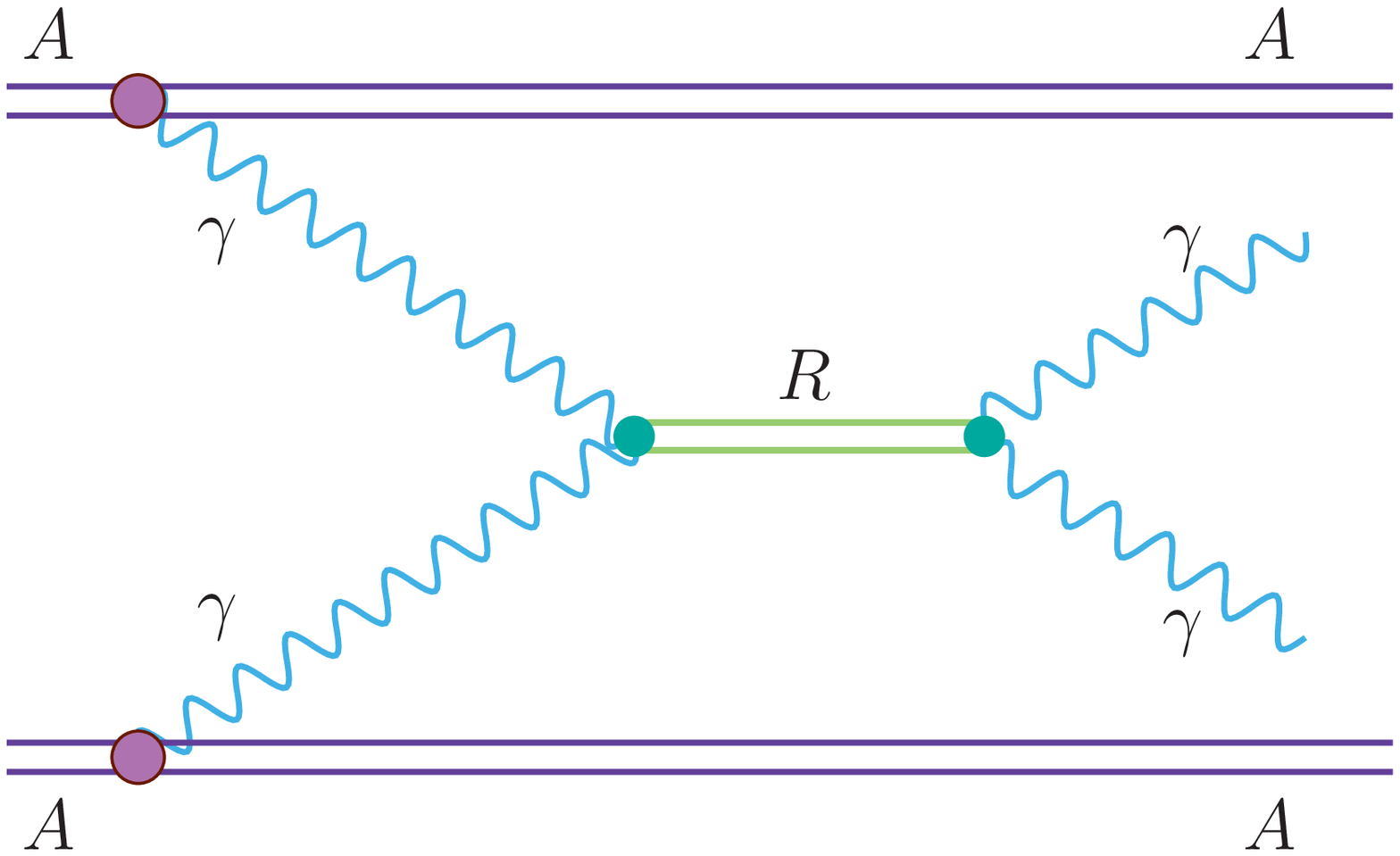}
	(c)\includegraphics[scale=0.22]{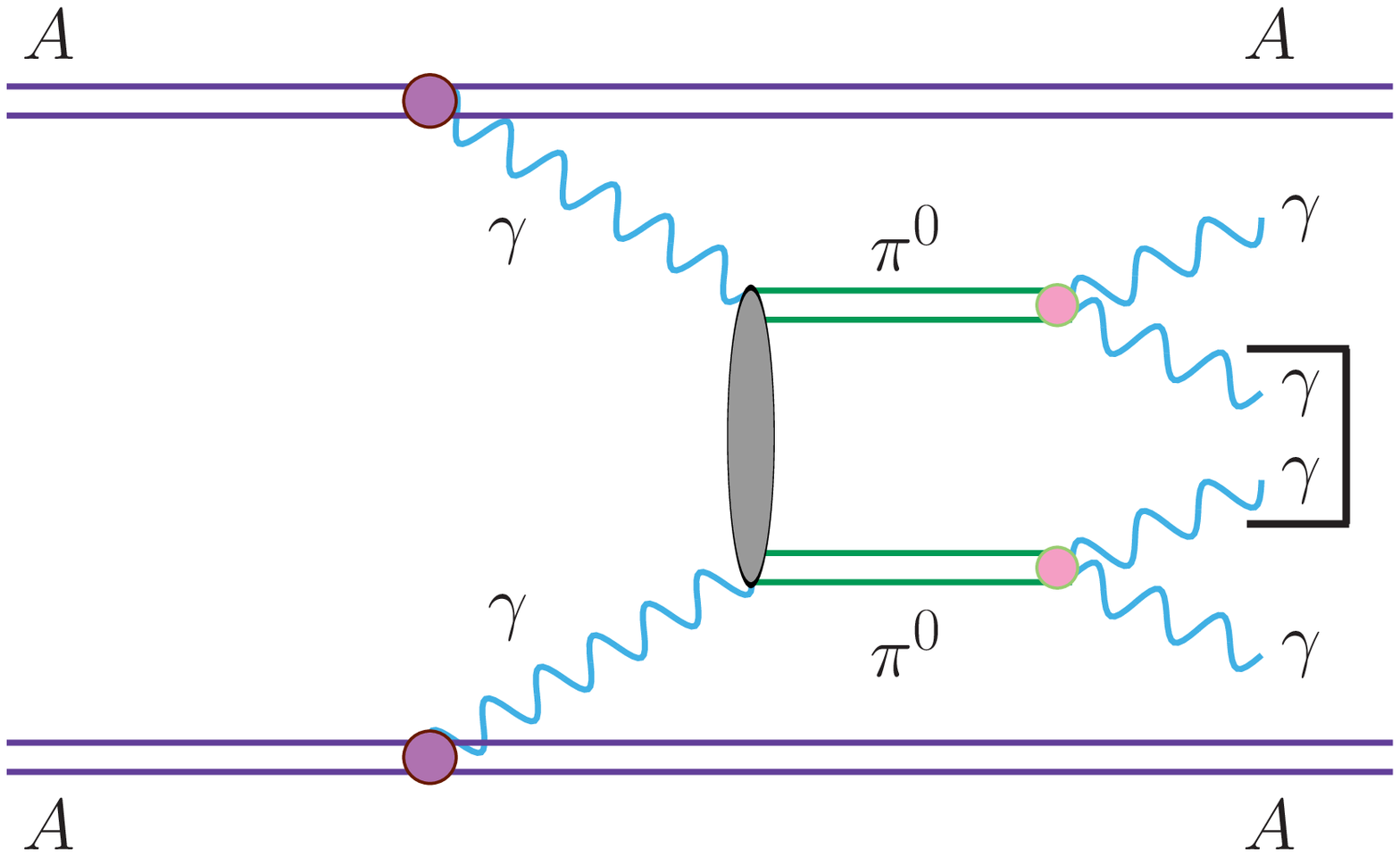}
	\caption{The continuum $\gamma \gamma \to \gamma \gamma$ scattering (a) 
		which includes box diagrams with leptons and quarks only \cite{Klusek-Gawenda:2016euz}.
		Panel (b) shows diagram for $s$-channel $\gamma\gamma \to $ pseudoscalar/scalar/tensor 
		resonances which contribute to the $\gamma \gamma \to \gamma \gamma$ process.
		Diagram (c) presents 
		background related to the~$\gamma\gamma \to \pi^0\pi^0$
		subprocess when only one photon from each $\pi^0 \to \gamma\gamma$
		decay is measured. }
	\label{fig:diagrams}       
\end{figure}
\section{Theoretical predictions}
\label{sec-2}
Studies are dedicated to ALICE and LHCb detector. The limitation of the 
ALICE detector is given by the detector geometry: 
$|\eta_{\gamma}|$ < 0.9  and for LHCb we have: 2.0 < $\eta_{\gamma}$ < 4.5.
Photons outside these regions are undetected. 
The ALICE experiment can measure photons for energies $E_{\gamma}$ > 0.2 GeV
and LHCb: $E_{t,\gamma}$ > 0.2 GeV (taking into account $m_\gamma$ = 0
one gets $E_{t,\gamma} = p_{t,\gamma}$).
The calculations include effect of experimental energy resolution.
For ALICE \cite{ALICE_thesis} and LHCb conditions \cite{Govorkova:2015vqa} the resolution functions are parametrized~as:
\begin{equation}
\left(\frac{\sigma_{E_\gamma}}{E_\gamma} \right)^{ALICE} =  1 \% \;, 
\qquad
\left(\frac{\sigma_{E_\gamma}}{E_\gamma} \right)^{LHCb} =   \frac{0.085}{\sqrt{E_\gamma}} + \frac{0.003}{E_\gamma}+ 0.008 \, .
\label{energy_resolution}
\end{equation}
In purely theoretical calculation
both signal photons have the same transverse momentum values $p_{t,\gamma}$
and transverse momentum of the diphoton pair is the Dirac-delta function $\delta (p_{t,\gamma\gamma})$.
In order to include more realistic conditions, we can use Gaussian distribution 
to obtain a~smearing in invariant mass or in scalar asymmetry 
($A_S = (|\vec{p_{t1}}|-|\vec{p_{t2}}|)/(|\vec{p_{t1}}|+|\vec{p_{t2}}|)$ ).
Then the transverse momenta of each of the photons takes the form:
$p_{i,t} = p_t + \left(\frac{p_t}{E_i} \right) \delta E_i$.
%
%
Here $\delta E_{i=1,2}$ is random number with Gaussian distribution 
with $\sigma_{E_i}$ given by Eq. (\ref{energy_resolution}).

\begin{figure}[h]
	\centering
	\includegraphics[scale=0.25]{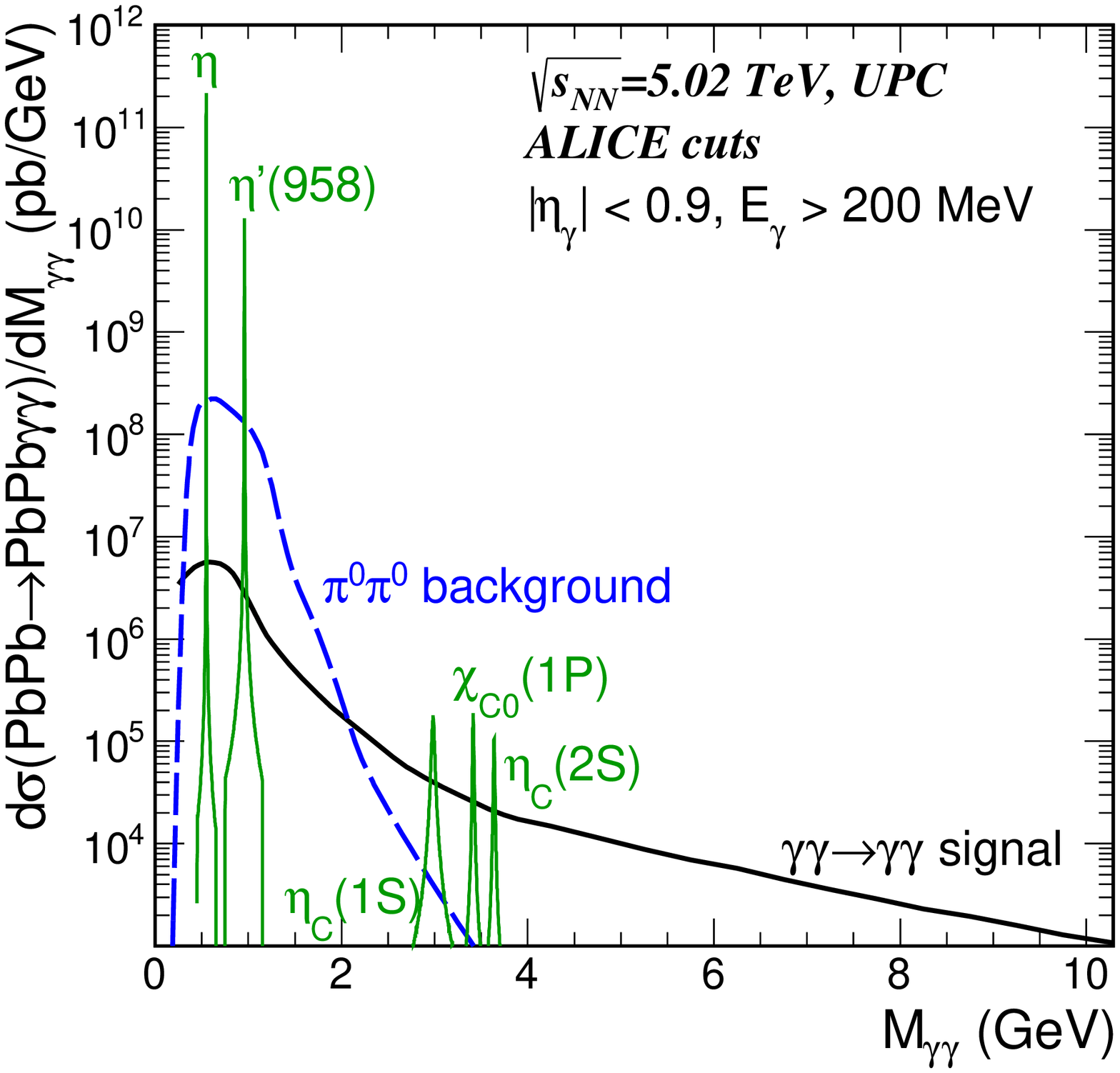} 
	\includegraphics[scale=0.25]{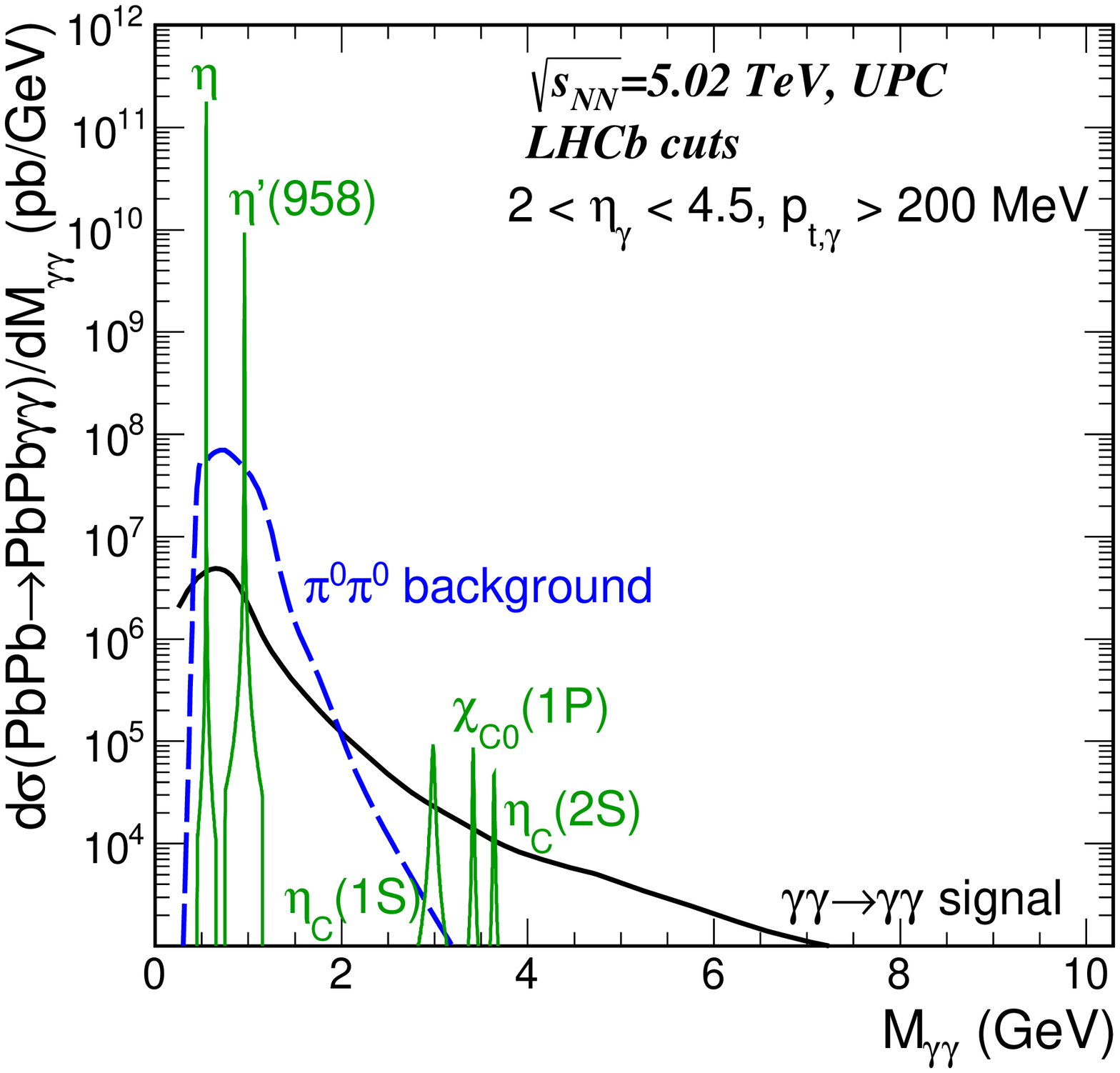}
	\caption{Diphoton invariant mass distribution for kinematical cuts 
		which are given in the figure:
		left panel is for ALICE experiment
		and right panel corresponds to LHCb experiment.
		The black solid lines are the signal due to 
		Standard Model box contribution, the blue dashed lines correspond to
		the $\pi^0\pi^0$ background, while solid green lines are for
		resonant mesonic states.}
	\label{fig:dsig_dW}       
\end{figure}

Fig.~\ref{fig:dsig_dW} shows comparison of contributions 
which come from the continuum $\gamma\gamma \to \gamma\gamma$ scattering
and $\gamma\gamma \to$ resonances $\to \gamma \gamma$
as well as the background. The background means the cases when
only two (out of four) photons pass the fiducial requirements.
This contribution dominates at low invariant diphoton masses (< 2 GeV).
These figures suggest that one could be able to measure the 
$\gamma \gamma \to \gamma \gamma$ scattering above $W_{\gamma\gamma}$ > 2 GeV.
In the case of this 1-dimensional distribution,
including experimental energy resolution modifies 
resonance signal.
Total cross section is of course still the same
but smearing of a resonance produces a peak about one order of magnitude
smaller than without experimental resolution.
The included energy resolution has a significance mainly at $\eta$ and $\eta'(958)$
peaks.

\begin{figure}[h]
	\centering
	\includegraphics[scale=0.25]{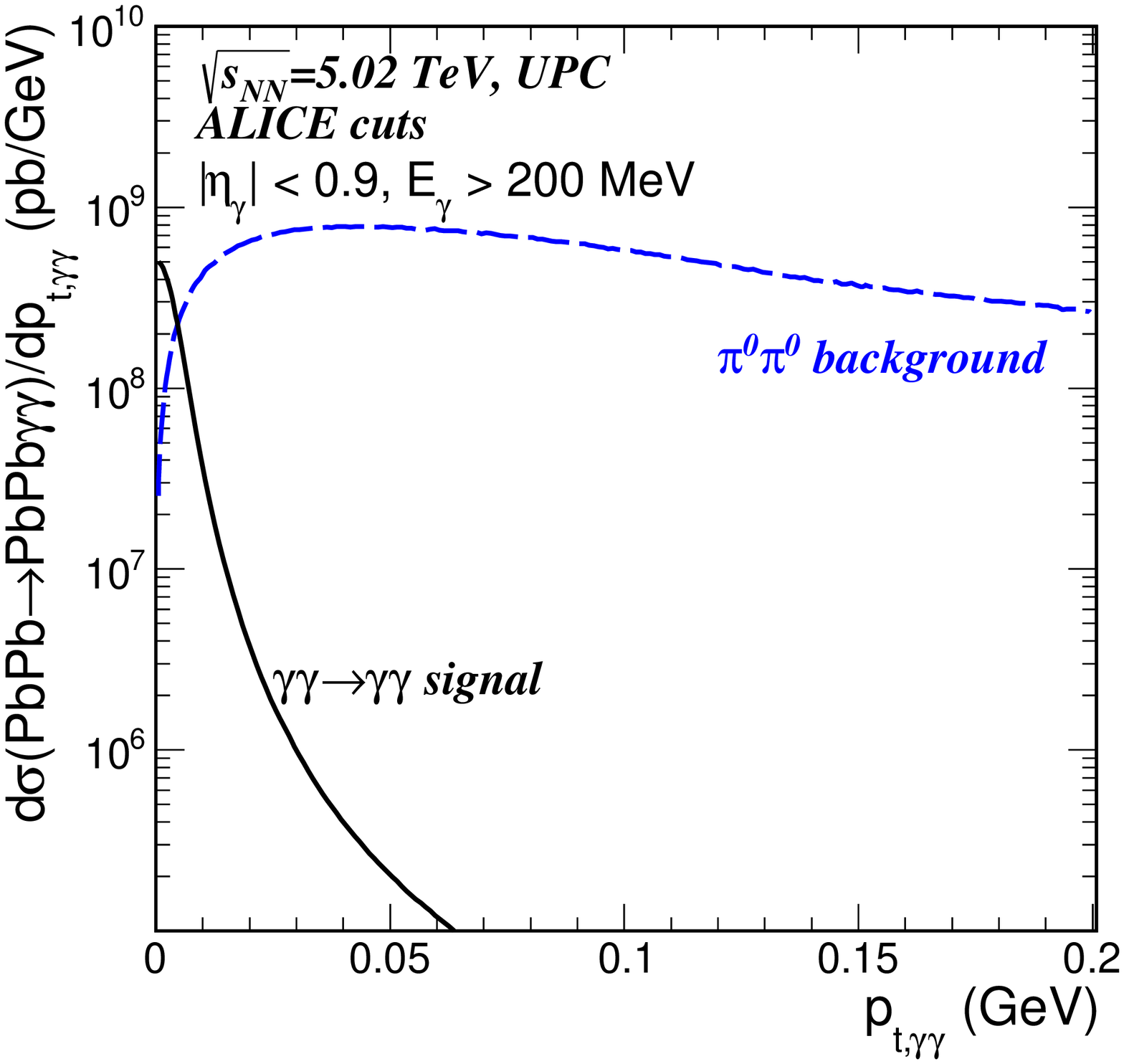}
	\includegraphics[scale=0.25]{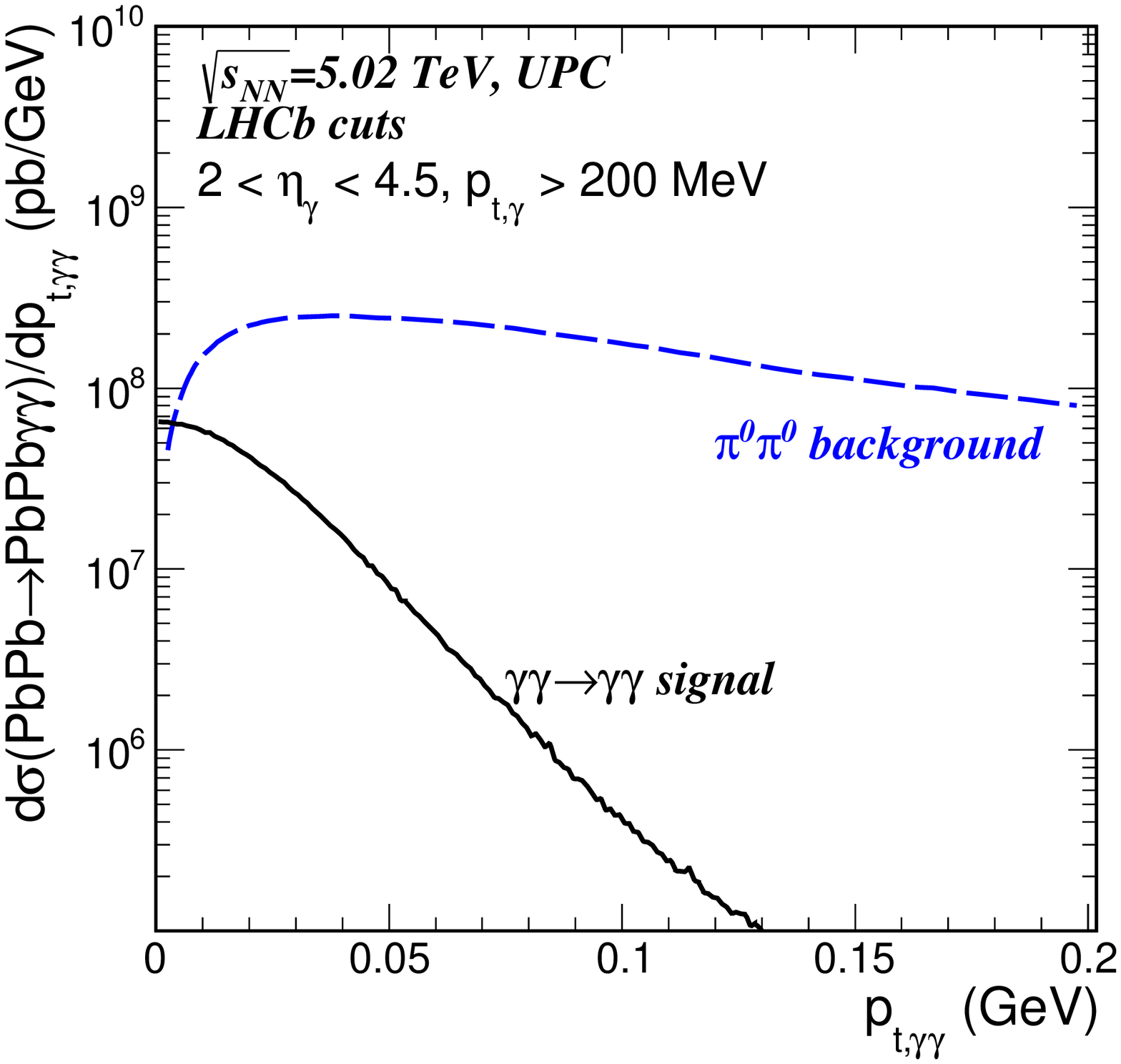}
	\caption{The differential cross section as a function of $p_{t,\gamma\gamma} =  (|\vec{p_{t1}} + \vec{p_{t2}}|)$ (transverse momentum of the diphoton pair)
	for the PbPb$\to$PbPb$\gamma\gamma$ reaction at $\sqrt{s_{NN}}=5.02$ TeV. }
	\label{fig:dsig_dptsum_bkg_and_signal}
\end{figure}

The measurement of two out of four photons for the $\pi^0\pi^0$ signal
leads to relatively large transverse momenta of the measured $\gamma\gamma$ pair
as shown in Fig.~\ref{fig:dsig_dptsum_bkg_and_signal}.
The left panel corresponds to the ALICE kinematical limitation 
and the right panel shows distribution for the LHCb cut.
Theoretically, fermionic box signal takes a form of
the Dirac-delta function, but due to experimental energy resolution,
one can observe some smearing and one can see a very limited region of 
very small $p_{t,\gamma\gamma} =  (|\vec{p_{t1}} + \vec{p_{t2}}|)$ 
where the $\gamma\gamma \to \gamma \gamma$ signal is above the $\pi^0\pi^0$ background.
It seems more difficult 
to separate the background from the signal for the LHCb cuts (right panel). 
The situation for the ALICE 
fiducial region (left panel) looks a~little bit better.  
By imposing extra cuts on $p_{t,\gamma\gamma}<$ 0.005, 0.01 and 0.02 GeV
one can reduce the ''unwanted'' background \cite{KGMSSz_2018}, 
however no complete removing of the background is possible.
The cut on $p_{t,\gamma\gamma}$ seems the most
efficient to reduce the $\pi^0\pi^0$ background.

\begin{table}[!htb]
	\centering
	\caption{Total nuclear cross section in nb.
		Only ''standard'' cuts for ALICE and LHCb are imposed.}
	\label{tab:events}
	\begin{tabular}{lrrrr}
		\hline
		Energy   &  \multicolumn{2}{c}{$W_{\gamma\gamma} = (0-2)$ GeV}  
		&  \multicolumn{2}{c}{$W_{\gamma\gamma}>$ 2 GeV}   \\ \hline
		Fiducial region	& ALICE		& LHCb		& ALICE		& LHCb	\\ \hline 
		boxes       &   4 890 	& 3 818		& 146		& 79  	\\ 
		$\pi^0\pi^0$ background  & 135 300	& 40 866	& 46 		& 24	\\ 
		$\eta$		& 722 573	& 568 499	& 			&		\\ 
		$\eta'(958)$&  54 241	&  40 482	& 			&		\\ 
		$\eta_c(1S)$&			&			& 9			& 5		\\ 
		$\chi_{c0}(1P)$& 		&			& 4			& 2		\\ 
		$\eta_c(2S)$&			&			& 2			& 1		\\ \hline	 
	\end{tabular}
\end{table}

The concluding section includes a Table with total cross section 
for ultraperipheral lead-lead collision at the LHC.
As can be seen,
the largest cross section is obtained for $\eta$ resonant production.
The Table~\ref{tab:events} includes a list of total nuclear cross sections 
in two ranges of invariant mass:
the first one is for the value of invariant mass from 0 to 2 GeV,
second one for $M_{\gamma\gamma}$ larger than 2 GeV.
Here $M_{\gamma\gamma}^{max}=$ 50 GeV for fermionic boxes,
$M_{\gamma\gamma}^{max}=$ 5 GeV for the $\pi^0\pi^0$ background
and $M_{\gamma\gamma}= (m_R - 1 \mbox{ GeV}, m_R + 1 \mbox{ GeV})$ for resonances
were assumed.
The background contribution, as was discussed above,
can be reduced. Here no extra cuts on the background are imposed.
Due to large masses of $\eta_c(1S)$, $\chi_{c0}(1P)$ and $\eta_c(2S)$ mesons,
the contribution from these resonances appears only at ''higher'' energies.
These cross sections are, however, very small in comparison to the other mechanisms.
The $\gamma\gamma$ scattering through $\eta$ and $\eta'$ resonances 
should be easily measurable. A precise measurement of the continuum part of the signal  for $W_{\gamma\gamma}<$ 2 GeV is not completely clear and requires 
experimental studies.

\begin{acknowledgement}
This work has been supported by the Polish National Science Center
grant DEC-2014/15/B/ST2/02528. 
\end{acknowledgement}
%
%
%

\end{document}